\documentclass[showpacs,showkeys,floatfix,prb,twocolumn]{revtex4}
\usepackage{graphicx}
\usepackage{dcolumn}
\usepackage{bm}
\usepackage{hyperref}
\def\vecr{{\bf r}}
\def\vece{{\bf e}}

\begin{document}

\title{Structure of metastable 2D liquid helium}
\author{V. Apaja} 
\affiliation{
Department of Physics, P.O. Box 35, FIN-40014, University of Jyv\"askyl\"a, Finland} 
\author{M. Saarela}
\affiliation{Department of Physical Sciences, P.O. Box 3000, FIN-90014,
 University of Oulu, Finland}

\begin{abstract}
We present diffusion Monte Carlo (DMC) results on a novel metastable,
superfluid phase in two-dimensional $^4$He at densities higher than
0.065\AA $^{-2}$.  The state is above the crystal ground state in
energy and it has anisotropic, hexatic orbital order. This implies
that the liquid--solid phase transition has two stages: A second order
phase transition from the isotropic superfluid to the hexatic
superfluid, followed by a first order transition that localizes atoms
into the triangular crystal order. This metastable hexatic phase
offers a natural explanation for the superflow in the supersolid
$^4$He and the possibility of a Kosterlitz-Thouless type phase
transition with increasing temperature.
\end{abstract}

\maketitle

Metastable states are transient, excited states that have a relatively
long lifetime, and may appear in the absence of an external
disturbance that would trigger the transition to the ground state.
Three--dimensional helium has a metastable high--density (pressure)
liquid phase, observed in laboratory experiments by Ishiguro {\em et
al.}\cite{ishiguro-caupin-balibar-06} and Werner {\em et
al.}\cite{balibar_160bars_04}. The liquid phase is measured up to 160
bar - far above the liquid solid freezing pressure of 25.3 bar. Also
Pearce {\em et al.} \cite{pearce-etal-2004} observed metastable liquid
at pressures up to 40 bar in helium immersed in gelsil pores. These
types of metastable states are typical for a first order phase
transition where latent heat must be released to make the transition
from liquid to solid phase. Diffusion Monte Carlo (DMC) simulations by
Vranjes {\em et al.}\cite{vranjes-boronat-casulleras-cazorla-05}
confirmed that a metastable state is superfluid with a finite
condensate fraction and has a roton minimum in the excitation spectrum
up to 275 bar, but no upper limit to this behavior is proposed.
 
Variational calculations of both 2D and 3D helium liquid suggest that
the isotropic low--density liquid state becomes unstable against
formation of an anisotropic liquid state before the solidification
pressure is reached
\cite{halinen-apaja-gernoth-saarela-00,apaja-halinen-saarela-00}. This
phase transition is of second order. No latent heat is required in the
transition and thus it can not support metastable states. In classical
fluids the corresponding anisotropic phase is named hexatic phase
after the proposal made by Halperin and
Nelson\cite{Halperin-Nelson-PRB}. Up to now, very large scale
simulations \cite{jaster-99,mak-06} have been performed with a simple
two-dimensional hard disk fluid to verify the theory of the continuous
phase transition, where hexatic phase is the intermediate phase before
the full solid order. These results seem to point toward a weakly
first order phase transition. 
  
Observation of the nonclassical rotational inertia (NCRI) by Kim and
Chan\cite{kim-chan-04,kim-chan-06} challenged our understanding of the
solid $^4$He phase. Since then both experimental and theoretical
results seem to conclude that a perfect crystal can not be superfluid
\cite{Chan2008,Prokofev2007,boninsegni-06,rittner:165301}. A strong
dependence of the superfluid fraction on crystal annealing supports
the idea that some kind of a metastable state could be responsible of
these observation. Boninsegni at al.\cite{boninsegni-06} proposed a
glassy phase, but recent experiments on the specific heat at very low
temperatures cast some doubts on that proposal
\cite{LinClarkChan2007}.  Recently Sasaki et al. \cite{sasaki-science}
have shown that the transport of mass in the supersolid $^4$He can
take place along grain boundaries. It requires that boundary layers
form a quasi two-dimensional superfluid. The puzzle is that superfluid
fraction does not scale with the amount of grain boundaries in the
sample. Nevertheless, numerical simulations have shown superflow at
grain boundaries \cite{pollet:135301}.
The superfluid fraction seems to support vortices, which
collect $^3$He impurities \cite{Anderson2007,kim:065301}, and the
phase transition to supersolid phase could then be related to the
Kosterlitz-Thouless type phase transition. Unusual behavior of the
shear modulus in solid $^4$He has also been observed
\cite{DayBeamish2007} at the same temperature range where the
supersolidity appears in the torsional oscillator measurements.

In this article we present results on DMC simulation of the
two--dimensional $^4$He at zero temperature and show that the hexatic,
high-pressure metastable liquid phase exists slightly above the
crystal phase in energy. The phase transition from liquid to solid has
two stages when the pressure increases. It is triggered by the second
order transition from the liquid to the hexatic phase, but then the
first order transition to the solid order requires external
perturbation.

The ground state properties of zero temperature 2D $^4$He have been
studied using Monte Carlo methods by Giorgini {\it et
  al.}\cite{giorgini-boronat-casulleras-96} and by Whitlock {\it et
  al.}\cite{whitlock-chester-kalos-88}, who find that the ground state
at high densities is a triangular solid phase. Helium layers on a
substrate, such as graphite, and in porous media have been thoroughly
studied using theory
\cite{clements-epstein-krotscheck-saarela-93,whitlock-chester-khrisnamachari-98,apaja-krotscheck-03b}
and experiments \cite{lauter-godfrin-frank-leiderer-92}.  The full
phase diagram at finite temperature has been calculated using path
integral Monte Carlo by Gordillo and Ceperley
\cite{gordillo-ceperley-98}. Also the change in the angular order in
the liquid--solid transition has been discussed using the variational
shadow wave function \cite{krishnamachari-chester}.  Our new result
complements those properties with a metastable superfluid state, which
has the hexatic two-particle structure.

We first describe how structural properties were obtained from the
simulation. Diffusion Monte Carlo
\cite{reynolds-ceperley-alder-lester-82} is a zero--temperature method
that uses a large number (in this work $\sim 500-1000$) of independent
$N$--atom simulations, {\em walkers}, to statistically represent an
imaginary time (marked $\tau$) evolution process to a wave function
$\Psi(\tau)$. In principle this projects out the excited states and
one proceeds to sample the properties of the ground state $\phi_0$. In
a simulation metastability arises if it is improbable that one sees
the asymptotic result $\Psi(\tau)\to \phi_0$ within the limited
simulation time. In other words, there may only be a very narrow
random--walk path that takes the evolution to the ground
state. Statistics is always improved with importance sampling and a
chosen trial state. With importance sampling one biases the random
walk, usually to the effect that the ground state is reached
faster. For example, if one imposes a triangular solid symmetry to the
trial state one obtains the properties of the premeditated solid
phase.\cite{whitlock-chester-kalos-88}
  
For helium we use the McMillan trial wave function, modulated with 
an angular component,
\begin{equation}
\varphi_T = \prod_{i<j}\exp\left[\alpha \frac{\cos(m\phi_{ij})-1}{r_{ij}}\right] 
\exp\left[-\left(\frac{r_{ij}}{\beta}\right)^\mu\right]\;.
\label{eq:trial}
\end{equation}
The variational constants are $\beta \sim 3.3$~\AA\ and $\mu=4$.  Here
$m$ is an even integer and the angle between atoms at $\vecr_i$ and
$\vecr_j$ is defined $\cos(\phi_{ij}) =
\frac{\vecr_j-\vecr_i}{r_{ij}}\cdot\hat\vece_0$, with respect to a
reference direction $\hat \vece_0$ and $r_{ij}=|\vecr_i-\vecr_j|$ is
the distance between the atoms $i$ and $j$.  The amplitude of the
trial angular structure is $\alpha \sim 0-0.6$.\cite{2D-comment1} This
wave function has no crystalline order and atoms are not fixed to
lattice sites, contrary to the Nosanow type solid trial state.
Also the anisotropic trial state is not the same as using, say, a substrate
potential, because the liquid structure cannot ignore the latter and a
potential induces a global order, and not just a local one, around
each atom.  Here we are not forcing anything upon the 2D
liquid itself, merely adding a possibility to measure the degree of
anisotropy from the simulation. Within statistical error, the energy
and the radial distribution function of the metastable state are
independent of the trial state.

The reference direction $\hat\vece_0$ is set externally, which enables
us to see if the high--density liquid orients itself to the direction
set by the trial wave function. In short, we estimate how
strongly the liquid binds to the globally defined orientation. From the
simulation we determine the pair distribution, expanded as
\begin{equation}
g({\bf r}) = \sum_{m=0}^\infty g_m(r) \exp{(im\theta)} \; ,
\label{eq:gmr}
\end{equation}
where $m$ is even and $\cos(\theta)=\hat\vecr\cdot \hat\vece_0$.  In
this work we keep terms up to $m=12$. We measure the pair distribution
rather than the momentum space static structure function, because we
expect to see only a very small effect and $g_m(r)$ can be accurately
deduced from the simulation.

If the trial state is still a liquid, why should the angular part make
any difference? The angular part serves a dual purpose, in close
relation to the way quantities are measured in DMC. First, a
well--known fact is that if the trial state is not too far from the
eigenstate $\phi$, one can approximate the expectation values of
operator ${\cal A}$ by the so--called {\em extrapolated estimator},
$\langle\phi |{\cal A}|\phi\rangle \approx 2\langle \phi|{\cal
A}|\varphi_T \rangle - \langle \varphi_T| {\cal A}|\varphi_T \rangle$,
where $\langle\phi |{\cal A}|\phi\rangle$ is what we want to know,
$\langle \phi|{\cal A}|\varphi_T \rangle$ is what the average over
walkers gives, and $\langle \varphi_T| {\cal A}|\varphi_T \rangle$ is
the value in the trial state. This implies that even if there is no
angular order in $\phi$, the walkers simulate roughly half of the
order we put in the trial state $\varphi_T$. As a result the trial
state gently biases the walkers to favor a given globally oriented
local anisotropy. In this respect the trial state acts as a
perturbation. The extrapolation or the forward walking algorithm
discussed below removes this perturbation from the final expectation
values.

The second reason for using the angular part is that it enables us to
measure how this ``perturbation'' affects the high--density liquid in
{\em coordinate space}.  To see why, let us first assume that the
metastable liquid indeed has a global angular order with respect to
some fixed external (laboratory) direction, but we use a spherically
symmetric trial state. Then in the DMC simulation each walker picks up
a different, randomly chosen orientation, which, over a large number
of walkers, averages to zero. This is the reason why we need a
non--zero angular order in the trial state in order to see one in
$\phi$ when using a coordinate space observable. The fact that a trial
state, or importance sampling in general, can be used to actually make
a quantity observable in a Monte Carlo simulation is not frequently
mentioned in standard texts.

If available, we always use {\em unbiased estimators}, since
extrapolated estimators are known to give systematic error, especially
for structural quantities like the pair distribution. In this work we
apply the algorithm based on forward walking described in
Ref.~\onlinecite{casulleras-boronat-95}. One keeps track on how many
asymptotic off-springs a given walker will have, and weights the
present situation accordingly. Although the name indicates that these
results are unbiased by the trial state, this is not entirely true: as
discussed above, without the trial state angular structure there is no
chance of observing coordinate space angular order. In our case the
unbiased estimators are more like conditional results, valid for a
certain fixed amplitude of the angular term in the trial state, and we
note in passing that the unbiased and the extrapolated estimators
agree well. We have also checked that the period we follow the walkers
is long enough so that the asymptotic regime is reached, but
not too long to become unstable. Sarsa {\it et
  al.} \cite{sarsa-schmidt-magro-00} compared the pair structure of 3D
liquid helium obtained using the Path Integral Ground State (PIGS)
method and the forward walking DMC algorithm and find that the two
methods produce very accurately the same result.

\begin{figure}[tb]
\includegraphics[width=0.48\textwidth]{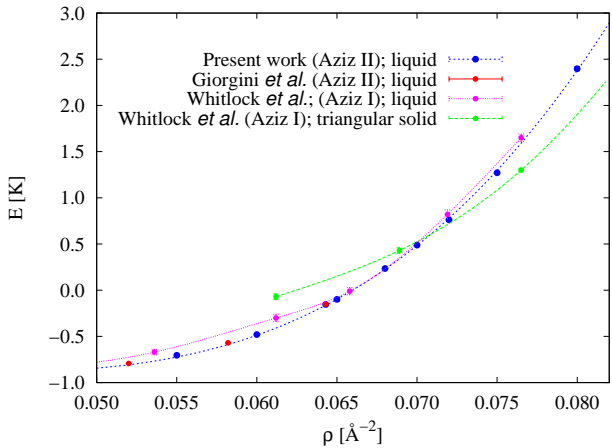}
\caption{The total energy of 2D helium as a function of density. The
  present data and the data by Giorgini {\em et
    al.}\cite{giorgini-boronat-casulleras-96} are computed using DMC
  with the revised Aziz HFDHE2 potential\cite{aziz-mccourt-wong-87}
  (``Aziz II'').  For comparison we show the energies of liquid and
  triangular solid by Whitlock {\em et
    al.}\cite{whitlock-chester-kalos-88}, who used the 1979 version of
  the Aziz HFDHE2 potential (``Aziz I'').\cite{aziz-nain-carley-79} }
\label{fig:energy}
\end{figure}

We have done simulations using a quadratic DMC algorithm, using mainly
64 or 120 atoms. Fig.~\ref{fig:energy} shows the total energy
vs. density in the high--density 2D liquid and triangular
solid. Notice the two slightly different He-He potentials used in the
QMC calculations. Our results agree well with the liquid energies
computed by Giorgini {\it et
  al.},\cite{giorgini-boronat-casulleras-96} available up to
$\rho<0.065$\AA$^{-2}$. For testing we also reproduced the Aziz-I
potential liquid energies reported by Whitlock {\it et
  al.}\cite{whitlock-chester-kalos-88}. Trial states with $\alpha$ in
the range 0-0.6 gave the same total energies within statistical error
bars. Also the spherically symmetric component of the radial
distribution function $g(r)\equiv g_{m=0}(r)$ is the same for any
trial state angular parameter $\alpha$.

\begin{figure}[bt]
\includegraphics[height=0.48\textwidth,angle=-90]{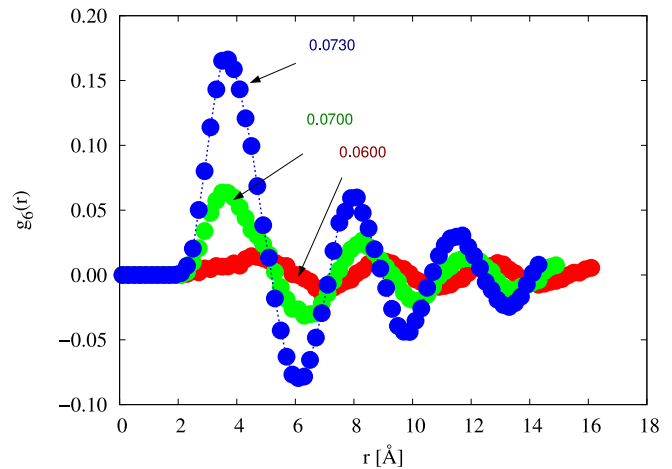}
\caption{The $g_{m=6}(r)$ component of the pair distribution function
  at densities $\rho=0.073$, 0.070, and 0.060\AA$^{-2}$, computed
  using the same $m=6$ trial state with $\alpha= 0.6$ and 64 atoms.
  The lowest density corresponds to a stable liquid and the two higher
  ones to the metastable liquid with a hexatic order.}
\label{fig:gmr}
\end{figure}

\begin{figure}[tb]
\includegraphics[width=0.48\textwidth]{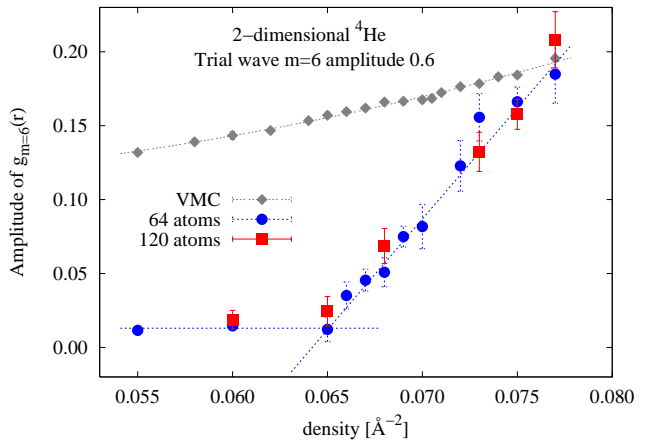}
\caption{The amplitude of the $g_{m=6}(r)$ component of the pair
  distribution as a function of density, computed using trial state
  with the amplitude $\alpha = 0.6$.  For reference, the points
  labeled ``VMC'' show the input trial state $g_{m=6}(r)$ amplitude.
  The lines are just guides to the eye.}
\label{fig:gmx}
\end{figure}

The phase transition from the isotropic to anisotropic liquid
generates angular dependence into the pair distribution function
$g({\bf r})$. The transition is continuous and the amplitude of the
angle dependence increases with increasing density. The angular
behavior is determined by the $m=6$ component of the expansion in
Eq. (\ref{eq:gmr}), which is in agreement with the point group
symmetry of the triangular lattice. Fig.~\ref{fig:gmr} shows the
component $g_{m=6}(r)$ at three densities near the freezing density,
computed using exactly the same trial state. While the stable liquid
($\rho=0.060$\AA$^{-2}$) is insensitive to the trial state, the
metastable liquid shows a clear externally oriented angular
structure. The amplitude of the short distance oscillations in
$g_{m=6}(r)$ increases with increasing density, but in long distances
these oscillations vanish, which means that the system remains in the
liquid state. The phase transition is made more apparent in
Fig.~\ref{fig:gmx}, where we plot the maximum value of $g_{m=6}(r)$
i.e. the amplitude of the first oscillation. We use the same trial
wave function at all densities and Fig.~\ref{fig:gmx} shows also the
``input amplitude'', computed using variational Monte Carlo
(VMC). While the trial wave function gives rise to angular amplitude
that slowly increases as the density increases, the DMC data shows a
clear onset of angular structure.  Above the density $0.065$\AA$^{-2}$
- remarkably close to the expected freezing density - the $g_{m=6}(r)$
component increases anomalously, yet much less than what would be
observed in freezing.\cite{2D-comment2} The fact that the DMC
algorithm reduces the amplitude from above 0.12 in the trial wave
function to less than 0.02 at low densities shows how little bias
there is, and that the forward walking DMC algorithm is indeed able to
remove the trial state angular structure if it is favorable.

\begin{figure}[hbt]
\includegraphics[width=0.48\textwidth]{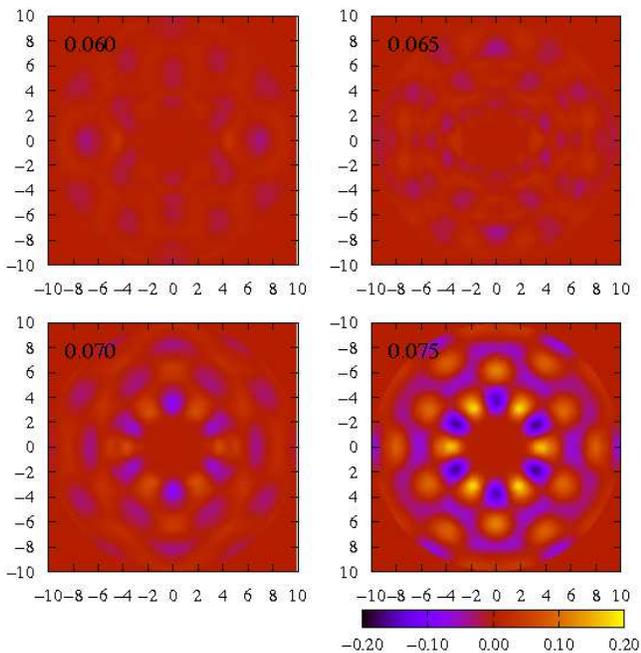}
\caption{The pair distribution $g({\bf r})$ summing $g_{m}(r)$ from
  $m=2$ to $m=12$ (without the radial component), as a function of
  density, computed using a trial state with $\alpha = 0.6$. Lighter
  areas correspond to positive (increased probability) and darker to
  negative values.  The labels in the figures show the densities in
  units \AA$^{-2}$.  The reference atom is in the middle and the we
  show a 20~\AA\ by 20~\AA\ box around it.  }
\label{fig:fullgr}
\end{figure}

In the metastable state the pair distribution function $g({\bf r})$
has a sixfold symmetry, depicted in Fig.~\ref{fig:fullgr}. To make the
modulation more visible we have subtracted the uninteresting radial
part $g(r)$. One might erroneously conclude that the apparent sixfold
symmetry seen in the pair distribution around {\em every} atom adds up
to having a triangular solid. However, a mere modulation in the
probability (relative to random) does not warrant that conclusion.
The full spatial crystal order requires releasing of latent heat
leading to the triangular structure in the {\it single particle
  density}\cite{gernoth-annals}. Here the single particle density
stays constant.

In the high---density, metastable $^4$He the angular
component $g_{m=6}(r)$  of the pair distribution function grows
above the freezing density, unlike any other component. To exclude the
possibility that this is an artifact due to the $m=6$ symmetry put
into the trial state, we repeated the calculation using a four--fold
symmetry in $\varphi_T$; in that case no anomalous increase in
$g_{m=4}(r)$ was observed. In our DMC calculations we used a
periodically repeated square box, which is not commensurate with a
triangular lattice and also won't enhance the sixfold symmetry
component. 
According to our results the metastable state is superfluid, although
the DMC method is not ideal for measuring the long range order in the
one-particle density matrix. We find no abrupt change in that long
range order when the density crosses the freezing density. The decay
of the long range tail remains very slow with increasing density as
one would expect from a 2D liquid $^4$He.

In conclusion we have shown that a novel metastable state in
two-dimensional $^4$He exists at high densities. It has the hexatic
orbital symmetry, but homogeneous single particle density. The
superfluidity of that state may explain the nonclassical rotational
inertia observed in supersolids.
As the state is metastable the crystal growth process determines
sensitively the fraction of $^4$He, which remains in the superfluid
phase. In rapid freezing of $^4$He within a narrow annular region by
Rittner and Reppy created a large superfluid fraction of a metastable
state up to 20 \% in the sample \cite{rittner:175302}. Annealing
removed the metastable state and the nonclassical rotational inertia
state almost completely disappeared \cite{rittner:165301}. We like to
draw attention to a similar metastable state in three dimensional
$^4$He where the ultrasound shock wave experiments pressurized $^4$He
far beyond the freezing density, and yet $^4$He remained superfluid.

We thank E. Krotscheck for many discussions and his hospitality during
our stay in the Johannes Kepler University in Linz, were part of the computations
were performed using the local computer resources. We are also
grateful to J. Boronat for stimulating discussions.

\bibliographystyle{apsrev}

\begin{thebibliography}{38}
\expandafter\ifx\csname natexlab\endcsname\relax\def\natexlab#1{#1}\fi
\expandafter\ifx\csname bibnamefont\endcsname\relax
  \def\bibnamefont#1{#1}\fi
\expandafter\ifx\csname bibfnamefont\endcsname\relax
  \def\bibfnamefont#1{#1}\fi
\expandafter\ifx\csname citenamefont\endcsname\relax
  \def\citenamefont#1{#1}\fi
\expandafter\ifx\csname url\endcsname\relax
  \def\url#1{\texttt{#1}}\fi
\expandafter\ifx\csname urlprefix\endcsname\relax\def\urlprefix{URL }\fi
\providecommand{\bibinfo}[2]{#2}
\providecommand{\eprint}[2][]{\url{#2}}

\bibitem[{\citenamefont{Ishiguro et~al.}(2006)\citenamefont{Ishiguro, Caupin,
  and Balibar}}]{ishiguro-caupin-balibar-06}
\bibinfo{author}{\bibfnamefont{R.}~\bibnamefont{Ishiguro}},
  \bibinfo{author}{\bibfnamefont{F.}~\bibnamefont{Caupin}}, \bibnamefont{and}
  \bibinfo{author}{\bibfnamefont{S.}~\bibnamefont{Balibar}},
  \bibinfo{journal}{Europ. Lett} \textbf{\bibinfo{volume}{75}},
  \bibinfo{pages}{91} (\bibinfo{year}{2006}).

\bibitem[{\citenamefont{Werner et~al.}(2004)\citenamefont{Werner, Beaume,
  Hobeika, Nascimbene, Herrman, Caupin, and Balibar}}]{balibar_160bars_04}
\bibinfo{author}{\bibfnamefont{F.}~\bibnamefont{Werner}},
  \bibinfo{author}{\bibfnamefont{G.}~\bibnamefont{Beaume}},
  \bibinfo{author}{\bibfnamefont{A.}~\bibnamefont{Hobeika}},
  \bibinfo{author}{\bibfnamefont{S.}~\bibnamefont{Nascimbene}},
  \bibinfo{author}{\bibfnamefont{C.}~\bibnamefont{Herrman}},
  \bibinfo{author}{\bibfnamefont{F.}~\bibnamefont{Caupin}}, \bibnamefont{and}
  \bibinfo{author}{\bibfnamefont{S.}~\bibnamefont{Balibar}},
  \bibinfo{journal}{J. Low Temp. Phys.} \textbf{\bibinfo{volume}{136}},
  \bibinfo{pages}{93} (\bibinfo{year}{2004}).

\bibitem[{\citenamefont{Pearce et~al.}(2004)\citenamefont{Pearce, Bossy,
  Schober, Glyde, Daughton, , and Mulders}}]{pearce-etal-2004}
\bibinfo{author}{\bibfnamefont{J.~V.} \bibnamefont{Pearce}},
  \bibinfo{author}{\bibfnamefont{J.}~\bibnamefont{Bossy}},
  \bibinfo{author}{\bibfnamefont{H.}~\bibnamefont{Schober}},
  \bibinfo{author}{\bibfnamefont{H.~R.} \bibnamefont{Glyde}},
  \bibinfo{author}{\bibfnamefont{D.~R.} \bibnamefont{Daughton}}, ,
  \bibnamefont{and} \bibinfo{author}{\bibfnamefont{N.}~\bibnamefont{Mulders}},
  \bibinfo{journal}{Phys. Rev. Lett.} \textbf{\bibinfo{volume}{93}},
  \bibinfo{eid}{145303} (\bibinfo{year}{2004}).

\bibitem[{\citenamefont{Vranje\v{s} et~al.}(2005)\citenamefont{Vranje\v{s},
  Boronat, Casulleras, and Cazorla}}]{vranjes-boronat-casulleras-cazorla-05}
\bibinfo{author}{\bibfnamefont{L.}~\bibnamefont{Vranje\v{s}}},
  \bibinfo{author}{\bibfnamefont{J.}~\bibnamefont{Boronat}},
  \bibinfo{author}{\bibfnamefont{J.}~\bibnamefont{Casulleras}},
  \bibnamefont{and} \bibinfo{author}{\bibfnamefont{C.}~\bibnamefont{Cazorla}},
  \bibinfo{journal}{Phys. Rev. Lett.} \textbf{\bibinfo{volume}{95}},
  \bibinfo{eid}{145302} (\bibinfo{year}{2005}).

\bibitem[{\citenamefont{Halinen et~al.}(2000)\citenamefont{Halinen, Apaja,
  Gernoth, and Saarela}}]{halinen-apaja-gernoth-saarela-00}
\bibinfo{author}{\bibfnamefont{J.}~\bibnamefont{Halinen}},
  \bibinfo{author}{\bibfnamefont{V.}~\bibnamefont{Apaja}},
  \bibinfo{author}{\bibfnamefont{K.~A.} \bibnamefont{Gernoth}},
  \bibnamefont{and} \bibinfo{author}{\bibfnamefont{M.}~\bibnamefont{Saarela}},
  \bibinfo{journal}{J. Low Temp. Phys.} \textbf{\bibinfo{volume}{121}},
  \bibinfo{pages}{531} (\bibinfo{year}{2000}).

\bibitem[{\citenamefont{Apaja et~al.}(2000)\citenamefont{Apaja, Halinen, and
  Saarela}}]{apaja-halinen-saarela-00}
\bibinfo{author}{\bibfnamefont{V.}~\bibnamefont{Apaja}},
  \bibinfo{author}{\bibfnamefont{J.}~\bibnamefont{Halinen}}, \bibnamefont{and}
  \bibinfo{author}{\bibfnamefont{M.}~\bibnamefont{Saarela}},
  \bibinfo{journal}{Physica B} \textbf{\bibinfo{volume}{284-288}},
  \bibinfo{pages}{29} (\bibinfo{year}{2000}).

\bibitem[{\citenamefont{Halperin and Nelson}(1979)}]{Halperin-Nelson-PRB}
\bibinfo{author}{\bibfnamefont{B.~I.} \bibnamefont{Halperin}} \bibnamefont{and}
  \bibinfo{author}{\bibfnamefont{D.~R.} \bibnamefont{Nelson}},
  \bibinfo{journal}{Phys. Rev. B} \textbf{\bibinfo{volume}{19}},
  \bibinfo{pages}{2457} (\bibinfo{year}{1979}).

\bibitem[{\citenamefont{Jaster}(1999)}]{jaster-99}
\bibinfo{author}{\bibfnamefont{A.}~\bibnamefont{Jaster}},
  \bibinfo{journal}{Phys. Rev. E} \textbf{\bibinfo{volume}{59}},
  \bibinfo{pages}{2594} (\bibinfo{year}{1999}).

\bibitem[{\citenamefont{Mak}(2006)}]{mak-06}
\bibinfo{author}{\bibfnamefont{C.~H.} \bibnamefont{Mak}},
  \bibinfo{journal}{Phys. Rev. E} \textbf{\bibinfo{volume}{73}},
  \bibinfo{eid}{065104} (\bibinfo{year}{2006}).

\bibitem[{\citenamefont{Kim and Chan}(2004)}]{kim-chan-04}
\bibinfo{author}{\bibfnamefont{E.}~\bibnamefont{Kim}} \bibnamefont{and}
  \bibinfo{author}{\bibfnamefont{M.~H.~W.} \bibnamefont{Chan}},
  \bibinfo{journal}{Nature} \textbf{\bibinfo{volume}{427}},
  \bibinfo{pages}{225} (\bibinfo{year}{2004}).

\bibitem[{\citenamefont{Kim and Chan}(2006)}]{kim-chan-06}
\bibinfo{author}{\bibfnamefont{E.}~\bibnamefont{Kim}} \bibnamefont{and}
  \bibinfo{author}{\bibfnamefont{M.~H.~W.} \bibnamefont{Chan}},
  \bibinfo{journal}{Phys. Rev. Lett.} \textbf{\bibinfo{volume}{97}},
  \bibinfo{pages}{115302} (\bibinfo{year}{2006}).

\bibitem[{\citenamefont{Chan}(2008)}]{Chan2008}
\bibinfo{author}{\bibfnamefont{M.~H.~W.} \bibnamefont{Chan}},
  \bibinfo{journal}{Science} \textbf{\bibinfo{volume}{319}},
  \bibinfo{pages}{1207} (\bibinfo{year}{2008}).

\bibitem[{\citenamefont{Prokof'ev}(2007)}]{Prokofev2007}
\bibinfo{author}{\bibfnamefont{N.~V.} \bibnamefont{Prokof'ev}},
  \bibinfo{journal}{Adv. Phys.} \textbf{\bibinfo{volume}{56}},
  \bibinfo{pages}{381} (\bibinfo{year}{2007}).

\bibitem[{\citenamefont{Boninsegni et~al.}(2006)\citenamefont{Boninsegni,
  Prokof'ev, and Svistunov}}]{boninsegni-06}
\bibinfo{author}{\bibfnamefont{M.}~\bibnamefont{Boninsegni}},
  \bibinfo{author}{\bibfnamefont{N.}~\bibnamefont{Prokof'ev}},
  \bibnamefont{and}
  \bibinfo{author}{\bibfnamefont{B.}~\bibnamefont{Svistunov}},
  \bibinfo{journal}{Phys. Rev. Lett.} \textbf{\bibinfo{volume}{96}},
  \bibinfo{eid}{105301} (\bibinfo{year}{2006}).

\bibitem[{\citenamefont{Rittner and Reppy}(2006)}]{rittner:165301}
\bibinfo{author}{\bibfnamefont{A.~S.~C.} \bibnamefont{Rittner}}
  \bibnamefont{and} \bibinfo{author}{\bibfnamefont{J.~D.} \bibnamefont{Reppy}},
  \bibinfo{journal}{Phys. Rev. Lett.} \textbf{\bibinfo{volume}{97}},
  \bibinfo{eid}{165301} (\bibinfo{year}{2006}).

\bibitem[{\citenamefont{Lin et~al.}(2007)\citenamefont{Lin, Clark, and
  Chan}}]{LinClarkChan2007}
\bibinfo{author}{\bibfnamefont{X.}~\bibnamefont{Lin}},
  \bibinfo{author}{\bibfnamefont{A.~C.} \bibnamefont{Clark}}, \bibnamefont{and}
  \bibinfo{author}{\bibfnamefont{M.~H.~W.} \bibnamefont{Chan}},
  \bibinfo{journal}{Nature} \textbf{\bibinfo{volume}{449}},
  \bibinfo{pages}{1025} (\bibinfo{year}{2007}).

\bibitem[{\citenamefont{Sasaki et~al.}(2006)\citenamefont{Sasaki, Ishiguro,
  Caupin, Maris, and Balibar}}]{sasaki-science}
\bibinfo{author}{\bibfnamefont{S.}~\bibnamefont{Sasaki}},
  \bibinfo{author}{\bibfnamefont{R.}~\bibnamefont{Ishiguro}},
  \bibinfo{author}{\bibfnamefont{F.}~\bibnamefont{Caupin}},
  \bibinfo{author}{\bibfnamefont{H.~J.} \bibnamefont{Maris}}, \bibnamefont{and}
  \bibinfo{author}{\bibfnamefont{S.}~\bibnamefont{Balibar}},
  \bibinfo{journal}{Science} \textbf{\bibinfo{volume}{313}},
  \bibinfo{pages}{1098} (\bibinfo{year}{2006}).

\bibitem[{\citenamefont{Pollet et~al.}(2007)\citenamefont{Pollet, Boninsegni,
  Kuklov, Prokof'ev, Svistunov, and Troyer}}]{pollet:135301}
\bibinfo{author}{\bibfnamefont{L.}~\bibnamefont{Pollet}},
  \bibinfo{author}{\bibfnamefont{M.}~\bibnamefont{Boninsegni}},
  \bibinfo{author}{\bibfnamefont{A.~B.} \bibnamefont{Kuklov}},
  \bibinfo{author}{\bibfnamefont{N.~V.} \bibnamefont{Prokof'ev}},
  \bibinfo{author}{\bibfnamefont{B.~V.} \bibnamefont{Svistunov}},
  \bibnamefont{and} \bibinfo{author}{\bibfnamefont{M.}~\bibnamefont{Troyer}},
  \bibinfo{journal}{Phys. Rev. Lett.} \textbf{\bibinfo{volume}{98}},
  \bibinfo{eid}{135301} (\bibinfo{year}{2007}).

\bibitem[{\citenamefont{Anderson}(2007)}]{Anderson2007}
\bibinfo{author}{\bibfnamefont{P.~W.} \bibnamefont{Anderson}},
  \bibinfo{journal}{Nature Physics} \textbf{\bibinfo{volume}{3}},
  \bibinfo{pages}{160} (\bibinfo{year}{2007}).

\bibitem[{\citenamefont{Kim et~al.}(2008)\citenamefont{Kim, Xia, West, Lin,
  Clark, and Chan}}]{kim:065301}
\bibinfo{author}{\bibfnamefont{E.}~\bibnamefont{Kim}},
  \bibinfo{author}{\bibfnamefont{J.~S.} \bibnamefont{Xia}},
  \bibinfo{author}{\bibfnamefont{J.~T.} \bibnamefont{West}},
  \bibinfo{author}{\bibfnamefont{X.}~\bibnamefont{Lin}},
  \bibinfo{author}{\bibfnamefont{A.~C.} \bibnamefont{Clark}}, \bibnamefont{and}
  \bibinfo{author}{\bibfnamefont{M.~H.~W.} \bibnamefont{Chan}},
  \bibinfo{journal}{Phys. Rev. Lett.} \textbf{\bibinfo{volume}{100}},
  \bibinfo{eid}{065301} (\bibinfo{year}{2008}).

\bibitem[{\citenamefont{Day and Beamish}(2007)}]{DayBeamish2007}
\bibinfo{author}{\bibfnamefont{J.}~\bibnamefont{Day}} \bibnamefont{and}
  \bibinfo{author}{\bibfnamefont{J.}~\bibnamefont{Beamish}},
  \bibinfo{journal}{Nature} \textbf{\bibinfo{volume}{450}},
  \bibinfo{pages}{853} (\bibinfo{year}{2007}).

\bibitem[{\citenamefont{Giorgini et~al.}(1996)\citenamefont{Giorgini, Boronat,
  and Casulleras}}]{giorgini-boronat-casulleras-96}
\bibinfo{author}{\bibfnamefont{S.}~\bibnamefont{Giorgini}},
  \bibinfo{author}{\bibfnamefont{J.}~\bibnamefont{Boronat}}, \bibnamefont{and}
  \bibinfo{author}{\bibfnamefont{J.}~\bibnamefont{Casulleras}},
  \bibinfo{journal}{Phys. Rev. B} \textbf{\bibinfo{volume}{54}},
  \bibinfo{pages}{6099} (\bibinfo{year}{1996}).

\bibitem[{\citenamefont{Whitlock et~al.}(1988)\citenamefont{Whitlock, Chester,
  and Kalos}}]{whitlock-chester-kalos-88}
\bibinfo{author}{\bibfnamefont{P.~A.} \bibnamefont{Whitlock}},
  \bibinfo{author}{\bibfnamefont{G.~V.} \bibnamefont{Chester}},
  \bibnamefont{and} \bibinfo{author}{\bibfnamefont{M.~H.} \bibnamefont{Kalos}},
  \bibinfo{journal}{Phys. Rev. B} \textbf{\bibinfo{volume}{38}},
  \bibinfo{pages}{2418} (\bibinfo{year}{1988}).

\bibitem[{\citenamefont{Clements et~al.}(1993)\citenamefont{Clements, Epstein,
  Krotscheck, and Saarela}}]{clements-epstein-krotscheck-saarela-93}
\bibinfo{author}{\bibfnamefont{B.~E.} \bibnamefont{Clements}},
  \bibinfo{author}{\bibfnamefont{J.~L.} \bibnamefont{Epstein}},
  \bibinfo{author}{\bibfnamefont{E.}~\bibnamefont{Krotscheck}},
  \bibnamefont{and} \bibinfo{author}{\bibfnamefont{M.}~\bibnamefont{Saarela}},
  \bibinfo{journal}{Phys. Rev. B} \textbf{\bibinfo{volume}{48}},
  \bibinfo{pages}{7450} (\bibinfo{year}{1993}).

\bibitem[{\citenamefont{Whitlock et~al.}(1998)\citenamefont{Whitlock, Chester,
  and Krishnamachari}}]{whitlock-chester-khrisnamachari-98}
\bibinfo{author}{\bibfnamefont{P.~A.} \bibnamefont{Whitlock}},
  \bibinfo{author}{\bibfnamefont{G.~V.} \bibnamefont{Chester}},
  \bibnamefont{and}
  \bibinfo{author}{\bibfnamefont{B.}~\bibnamefont{Krishnamachari}},
  \bibinfo{journal}{Phys. Rev. B} \textbf{\bibinfo{volume}{58}},
  \bibinfo{pages}{8704} (\bibinfo{year}{1998}).

\bibitem[{\citenamefont{Apaja and Krotscheck}(2003)}]{apaja-krotscheck-03b}
\bibinfo{author}{\bibfnamefont{V.}~\bibnamefont{Apaja}} \bibnamefont{and}
  \bibinfo{author}{\bibfnamefont{E.}~\bibnamefont{Krotscheck}},
  \bibinfo{journal}{Phys. Rev. Lett.} \textbf{\bibinfo{volume}{91}},
  \bibinfo{pages}{225302} (\bibinfo{year}{2003}).

\bibitem[{\citenamefont{Lauter et~al.}(1992)\citenamefont{Lauter, Godfrin,
  Frank, and Leiderer}}]{lauter-godfrin-frank-leiderer-92}
\bibinfo{author}{\bibfnamefont{H.~J.} \bibnamefont{Lauter}},
  \bibinfo{author}{\bibfnamefont{H.}~\bibnamefont{Godfrin}},
  \bibinfo{author}{\bibfnamefont{V.~L.~P.} \bibnamefont{Frank}},
  \bibnamefont{and} \bibinfo{author}{\bibfnamefont{P.}~\bibnamefont{Leiderer}},
  \bibinfo{journal}{Phys. Rev. Lett.} \textbf{\bibinfo{volume}{68}},
  \bibinfo{pages}{2484} (\bibinfo{year}{1992}).

\bibitem[{\citenamefont{Gordillo and Ceperley}(1998)}]{gordillo-ceperley-98}
\bibinfo{author}{\bibfnamefont{M.~C.} \bibnamefont{Gordillo}} \bibnamefont{and}
  \bibinfo{author}{\bibfnamefont{D.~M.} \bibnamefont{Ceperley}},
  \bibinfo{journal}{Phys. Rev. B} \textbf{\bibinfo{volume}{58}},
  \bibinfo{pages}{6447} (\bibinfo{year}{1998}).

\bibitem[{\citenamefont{Krishnamachari and
  Chester}(2000)}]{krishnamachari-chester}
\bibinfo{author}{\bibfnamefont{B.}~\bibnamefont{Krishnamachari}}
  \bibnamefont{and} \bibinfo{author}{\bibfnamefont{G.~V.}
  \bibnamefont{Chester}}, \bibinfo{journal}{Phys. Rev. B}
  \textbf{\bibinfo{volume}{61}}, \bibinfo{pages}{9677} (\bibinfo{year}{2000}).

\bibitem[{\citenamefont{Reynolds et~al.}(1982)\citenamefont{Reynolds, Ceperley,
  Alder, and Lester}}]{reynolds-ceperley-alder-lester-82}
\bibinfo{author}{\bibfnamefont{P.~J.} \bibnamefont{Reynolds}},
  \bibinfo{author}{\bibfnamefont{D.~M.} \bibnamefont{Ceperley}},
  \bibinfo{author}{\bibfnamefont{B.~J.} \bibnamefont{Alder}}, \bibnamefont{and}
  \bibinfo{author}{\bibfnamefont{W.~A.} \bibnamefont{Lester}},
  \bibinfo{journal}{J. Chem. Phys.} \textbf{\bibinfo{volume}{77}},
  \bibinfo{pages}{5593} (\bibinfo{year}{1982}).

\bibitem[{2D-({\natexlab{a}})}]{2D-comment1}
\bibinfo{note}{As Fig. 3 shows, states with larger angular order are
  variationally far from the final state, and one needs excessively many
  walkers to remove the statistical bias.}

\bibitem[{\citenamefont{Casulleras and Boronat}(1995)}]{casulleras-boronat-95}
\bibinfo{author}{\bibfnamefont{J.}~\bibnamefont{Casulleras}} \bibnamefont{and}
  \bibinfo{author}{\bibfnamefont{J.}~\bibnamefont{Boronat}},
  \bibinfo{journal}{Phys. Rev. B} \textbf{\bibinfo{volume}{52}},
  \bibinfo{pages}{3654} (\bibinfo{year}{1995}).

\bibitem[{\citenamefont{Sarsa et~al.}(2000)\citenamefont{Sarsa, Schmidt, and
  Magro}}]{sarsa-schmidt-magro-00}
\bibinfo{author}{\bibfnamefont{A.}~\bibnamefont{Sarsa}},
  \bibinfo{author}{\bibfnamefont{K.~E.} \bibnamefont{Schmidt}},
  \bibnamefont{and} \bibinfo{author}{\bibfnamefont{W.~R.} \bibnamefont{Magro}},
  \bibinfo{journal}{J. Chem. Phys.} \textbf{\bibinfo{volume}{113}},
  \bibinfo{pages}{1366} (\bibinfo{year}{2000}).

\bibitem[{\citenamefont{Aziz et~al.}(1979)\citenamefont{Aziz, McCourt, and
  Wong}}]{aziz-mccourt-wong-87}
\bibinfo{author}{\bibfnamefont{R.~A.} \bibnamefont{Aziz}},
  \bibinfo{author}{\bibfnamefont{F.~R.~W.} \bibnamefont{McCourt}},
  \bibnamefont{and} \bibinfo{author}{\bibfnamefont{C.~C.~K.}
  \bibnamefont{Wong}}, \bibinfo{journal}{J. Chem. Phys.}
  \textbf{\bibinfo{volume}{70}}, \bibinfo{pages}{4330} (\bibinfo{year}{1979}).

\bibitem[{\citenamefont{Aziz et~al.}(1987)\citenamefont{Aziz, Nain, Carley,
  Taylor, and McConville}}]{aziz-nain-carley-79}
\bibinfo{author}{\bibfnamefont{R.~A.} \bibnamefont{Aziz}},
  \bibinfo{author}{\bibfnamefont{V.~P.~S.} \bibnamefont{Nain}},
  \bibinfo{author}{\bibfnamefont{J.~C.} \bibnamefont{Carley}},
  \bibinfo{author}{\bibfnamefont{W.~J.} \bibnamefont{Taylor}},
  \bibnamefont{and} \bibinfo{author}{\bibfnamefont{G.~T.}
  \bibnamefont{McConville}}, \bibinfo{journal}{Mol. Phys.}
  \textbf{\bibinfo{volume}{61}}, \bibinfo{pages}{1487} (\bibinfo{year}{1987}).

\bibitem[{2D-({\natexlab{b}})}]{2D-comment2}
\bibinfo{note}{We also saw few cases of freezing with rapid increase in the
  angular order, similar to one reported in
  Ref.~\onlinecite{krishnamachari-chester}}.

\bibitem[{\citenamefont{Gernoth}(2000)}]{gernoth-annals}
\bibinfo{author}{\bibfnamefont{K.~A.} \bibnamefont{Gernoth}},
  \bibinfo{journal}{Ann. Phys.} \textbf{\bibinfo{volume}{61}},
  \bibinfo{pages}{285} (\bibinfo{year}{2000}).

\bibitem[{\citenamefont{Rittner and Reppy}(2007)}]{rittner:175302}
\bibinfo{author}{\bibfnamefont{A.~S.~C.} \bibnamefont{Rittner}}
  \bibnamefont{and} \bibinfo{author}{\bibfnamefont{J.~D.} \bibnamefont{Reppy}},
  \bibinfo{journal}{Phys. Rev. Lett.} \textbf{\bibinfo{volume}{98}},
  \bibinfo{eid}{175302} (\bibinfo{year}{2007}).

\end{thebibliography}

\end{document}